\documentclass[pre,amsmath,amssymb,floatfix,preprint,showpacs]{revtex4}
\usepackage{graphicx}
\usepackage{float}
\usepackage{bm}
\usepackage[utf8]{inputenc}

\begin{document}
\title{Phase behavior of the Confined Lebwohl-Lasher Model}
\author{N.G. Almarza, C. Mart\'{\i}n, and E. Lomba }
\affiliation{Instituto de Qu{\'\i}mica F{\'\i}sica Rocasolano, CSIC, Serrano 119, E-28006 Madrid, Spain }

\date{\today}
\begin{abstract}
The phase behavior of confined nematogens is studied
using the Lebwohl-Lasher model. For three dimensional systems the model
is known to exhibit a discontinuous nematic-isotropic phase transition,
whereas the corresponding two dimensional systems apparently show a continuous
Berezinskii-Kosterlitz-Thouless like transition.
In this paper we study the phase transitions of the Lebwohl-Lasher model when confined
between planar slits of different widths in order to establish the behavior of intermediate
situations between the pure planar model and the three-dimensional system, and compare with previous
estimates for the critical thickness, i.e. the slit width at which the
transition switches  from 
continuous to discontinuous.
\end{abstract}
\maketitle
\section{Introduction}

The Lebwohl-Lasher (LL)  model \cite{Lebwohl,Fabbri,Zhang,Priezjev,Luckhurst1,Luckhurst2}
is a lattice model of an anisotropic fluid. Each site of the lattice is occupied by a uniaxial molecule.
A molecule interacts exclusively with molecules located at its nearest-neighbor (NN) sites. The total potential
energy takes the form:
\begin{equation}
U = - \epsilon \sum_{<ij>} P_2( {\bf s}_i \cdot {\bf s}_j),
\end{equation}
where $\epsilon$ is the coupling parameter ($\epsilon > 0$), ${\bf s}_i$ and ${\bf s}_j$
are unit vectors that indicate the orientation of the molecules in the corresponding sites, $P_2$ is the
second degree Legendre polynomial, and $<ij>$ indicates that the sum is restricted to NN pairs
of sites. 
The LL model can be deemed as the lattice version of the hard sphere Maier-Saupe (HSMS) fluid 
\cite{MS1,MS2,Lomba1,Lomba2,Lomba3,Almarza1}. 
Most of the simulation work on the LL model has been carried out on simple cubic lattices for three-dimensional (3D)
systems, and square lattices for the two-dimensional (2D) case, although some variations have also been considered\cite{Luckhurst2}.

A number of papers have been devoted to the analysis of the phase diagram of the LL model
using computer simulation. The model in 3D  
has been found to exhibit
a discontinuous nematic-isotropic transition 
\cite{Luckhurst1,Fabbri,Zhang,Priezjev}. The planar Lebwohl-Lasher (PLL) model, defined on a square lattice
has also been treated extensively using computer 
simulation \cite{Kunz,Farinas-Sanchez,Chiccoli,Mondal,Paredes}. From this set of results
it has been suggested that the PLL model presents a topological
defect driven continuous transition of  
the Berezinskii-Kosterlitz-Thouless (BKT) type\cite{BKT1,BKT2}. Notice however, that some
differences between the transition of the PLL model and that of the two dimensional XY model (the
paradigm for the topological BKT behavior) have been recently reported\cite{Paredes}.

In this paper we will pay attention to the nature of the phase transitions
of this system under confinement in a slit pore, and will study the influence of
the pore width on the transition. Herein we will be dealing with slits
formed by neutral walls, by which the systems under consideration
will be, in fact, slab models. In this regard, rigorous results\cite{r02,r03,r04}
indicate that this type of models cannot support true long range order
at finite temperature (in common with bidimensional
systems\cite{nelson}). This implies that in our context of
confined/slab and planar systems, we may encounter phases with quasi-long
range orientational order, which will be here referred to as
quasi-nematics. From the point of view of simulation, the LL 
model confined in slit pores  was previously studied
by Cleaver and Allen \cite{cleaver}. They concluded that the system has
a critical thickness, $H_c$, below which there is no bulk-like transition.
The existence of such a multicritical point in the  $T-H$ plane (where $T$ is the temperature
and $H$ is the thickness of the slab) can be explained using
theoretical arguments\cite{Shnidman,telodagama,ferreira}. Nonetheless,
according to the theoretical approach of Telo da Gama and
Tarazona\cite{telodagama}, in the case of neutral walls, one should
expect $H_c\rightarrow\infty$. Here, 
we will address this issue  resorting to simulation techniques, and
analyzing the results  obtained for system sizes much larger 
than those considered in Ref.\cite{cleaver}. We will thus assess the
bounds proposed therein for such a possible critical thickness.

 In
close connection with this work, 
the effect of the confinement on the isotropic-nematic transition has
been studied in Ref.\cite{Almarza1} for the HSMS model, 
where it was found that for some temperatures the first order
isotropic-nematic transition can disappear when the system 
is confined in flat slits with thickness below a certain width $H_c(T)$. For smaller values of the pore width, $H < H_c(T)$, a
BKT-like transition appears. Nevertheless, in the HSMS model one has
to deal with density fluctuations that are not present
in the LL model, and this could influence the phase behavior. Notice, that it is however possible to introduce
density fluctuations within a lattice model, as it can be seen in the
so-called Lebwohl-Lasher lattice gas model\cite{Bates}. 

Also related to the present work, a very recent article by Fish and
Vink\cite{Fish} focuses on the effects of confinement on a generalized
version of the LL model, in which the angular dependent component of
the interaction is $\propto |{\bf s}_i\cdot {\bf s}_j|^p$. These authors
analyze the behavior of the model for values of $p \ge 8$ (note that
in the present instance $p=2$) for which
they show there is a well defined critical thickness that vanishes
for large values of $p$ when the phase transition becomes first order
even in the two-dimensional limit. 

In summary, when going from the LL bulk behavior to that of the
confined system, we should be able to sort out between various
possible scenarios. First, the transition between that isotropic and
nematic phase might be second order, being the ordered phase not
critical below the transition temperature, with a finite and non-zero
order parameter and a diverging susceptibility only at the critical
point. This situation is in principle ruled out by the exact results
that preclude the existence of long range order (i.e. a non vanishing
order parameter) in our model\cite{r02,r03,r04}. Another possibility,
would be the presence of a continuous BKT transition, in which below
the transition temperature the system exhibits quasi-long range
orientational order (quasi-nematic phase with a vanishing order
parameter) and the susceptibility diverges at all temperatures below
the transition temperature. Some subtle issues, regarding what a true
BKT transition implies in connection with the discussion of
Ref.~\cite{Paredes} will be addressed in later sections of this
paper. Finally, another alternative is illustrated by the generalized
XY and related models\cite{r11,r12,Domany,blotte,Vink07}, which for
sufficiently ``sharp and narrow'' interactions\cite{Vink07} have been
shown to undergo a first order transition between the isotropic and
quasi-nematic phases. It is thus, the aim of this work to provide
additional information in order to be able to discern between those
scenarios. 

The rest of the paper is organized as follows; after this
introduction, in section II we describe the simulation 
methodology and summarize the details of the calculations and systems
under consideration. In section III we present our main
results and discuss our most relevant conclusions.

\section{Simulation techniques}

We will deal with systems consisting of $L\times L \times H$
sites. Periodic boundary conditions (PBC) are applied on the $x-$ and $y-$
directions, and 
the systems are confined by neutral walls in the $z-$direction. For a
given slab thickness, $H$, 
results for different values of $L$  are taken into account in order
to perform the finite-size scaling analysis. We have here studied systems with
$H=1, 2, 3, 4, 5, 8$, and $16$. For each value of $H$ we have
considered a series of $L$ values, namely, 
$L=$ 10, 15, 20, 25, 30, 25, 40, 45, 50, 60, 80, 100, 120, 140, 160
and 200. 

In addition we have also simulated various systems using PBC in the
three spatial directions. In particular, 
systems with $H=16$ and different values of $L$, so as to analyze the
effects of the boundary conditions on the transitions. Fully cubic
systems $L\times L\times L$ with PBC were also simulated in order  to
represent the 3D bulk system. Obviously, we will not be dealing here
with ``true'' bulk systems, but we will use the results of
non-confined isotropic periodic systems, after a finite size scaling
analysis is performed, as a good approximation to the bulk system
results. For simplicity, these corrected results will be referred to
as ``bulk'' data. 

We have performed Monte Carlo simulations combining single particle
Monte Carlo steps with cluster algorithms\cite{Wolff,Priezjev,Lomba1} using multicluster moves\cite{Lomba0,Swendsen}. 
For given values of the system sizes, $L$, and $H$ we performed
independent simulation runs at several temperatures close to the range where the transitions are expected. The results
were analyzed using efficient re-weighting procedures \cite{Ferrenberg,Lomba0}. The simulation 
procedures have been adapted from our previous works 
to the simpler lattice system, and technical details can be found elsewhere \cite{Lomba1,Lomba2,Lomba3}.
In order to locate the isotropic-quasi-nematic transitions we monitored the
largest eigenvalue, $\lambda_+$ of Saupe's tensor\cite{Saupe}:
\begin{equation}
Q_{\alpha\beta} = \frac{1}{2 N} \sum_{i=1}^N \left( 3 s_{i}^{\alpha} s_i^{\beta} - \delta_{\alpha\beta} \right).
\end{equation}
For a given system size, described by the lengths $L$ and $H$ we can
define pseudo-critical temperatures, $T_c(L,H)$,   in terms of
 the behavior of $\lambda_+$ as a function of the temperature, and
 also in terms of the
 temperature dependence  of the susceptibility, this quantity being
 defined by means of the fluctuation of the order parameter as: 
\begin{equation}
\chi = N \left( <\lambda_+^2>-<\lambda_+>^2 \right)/k_B T.
\label{chi}
\end{equation}
In practice, we consider two criteria to determine the pseudocritical
points, namely the temperature at which $\chi$, as defined 
in Eq.(\ref{chi}), is maximum, and the temperature that gives the
largest value of $|d\lambda_+/dT|$. 
Then, one can use the pseudo-critical temperatures to extrapolate the
transition temperature in the thermodynamic 
limit ($L \rightarrow  \infty$). Following the usual practice \cite{Kunz,Mondal}, we have used both the
expected scaling for BKT transitions\cite{Kunz,Tomita},
\begin{equation}
T_c(L) = T_c + \frac{a_1}{(a_2 + \ln L)^2 },
\label{fit_bkt}
\end{equation}
and the scaling equation of second order transitions\cite{Kunz,Mondal}:
\begin{equation}
T_c(L) = T_c + a_1 L^{-1/\nu}.
\label{fit_ct}
\end{equation}
Notice that we have not used the loci of maxima of the excess heat capacity per
particle, $c_v$,   as an additional
criterion to define pseudo-critical temperatures. This alternative was
used in \cite{cleaver}, but most likely is not a good choice for
systems that might  exhibit a BKT-like transition (e.g. for small
values of $H$). In such a case,
the maximum in $c_v$ is not well defined and does not diverge with
increasing sample sizes. Therefore it is not obvious that its location
signals the presence of a phase transition. 
It is worth mentioning that we have implicitly assumed 
in Eq. (\ref{fit_bkt}) an exponential divergence of the
correlation length $\xi \propto e^{bt^{\nu}}$, with $\nu=1/2$;
this value of $\nu$ is known to be appropriate for the XY-model\cite{BKT2,Kunz,Tomita}.
We decided to use $\nu=1/2$ following Ref.(\onlinecite{Kunz}), due to the
fact that a sensible fitting of the simulation results to a non-linear
equation involving four adjustable parameters would require both
a much larger range of values of $L$ and very precise simulation results.

\section{Simulation Results}

\begin{figure}
\includegraphics[width=90mm,clip=]{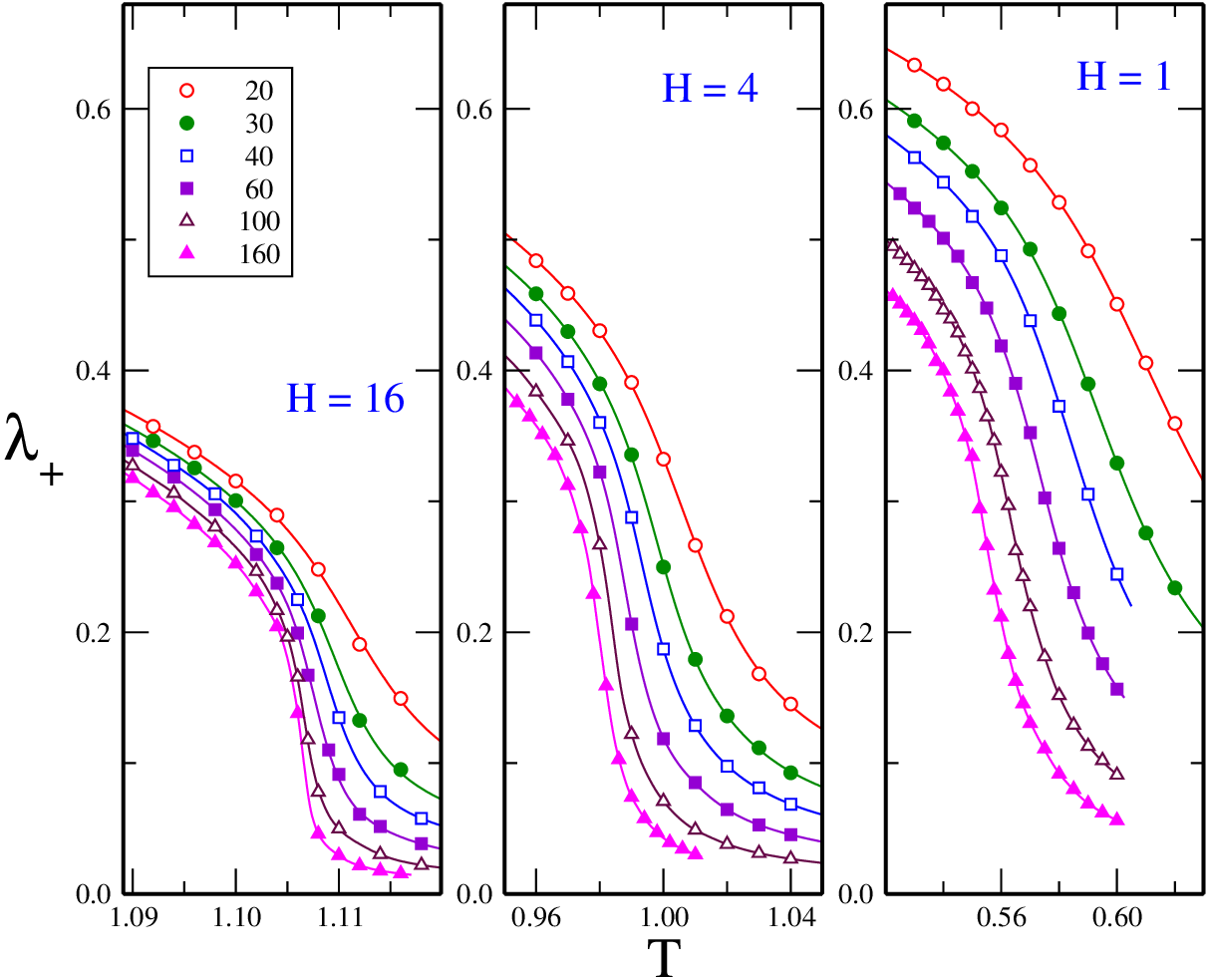}
\caption{(Color on line) Order parameter $\lambda_+$ as a function of the temperature for different system widths,
$H=1,4,16$ and different systems sizes, $L$. 
Symbols denote the result of different simulation runs, and lines
represent the results of the reweighting analysis. The legends in the figures indicate the
different values of $L$.}
\label{fig.lambda}
\end{figure}

\begin{figure}
\includegraphics[width=90mm,clip=]{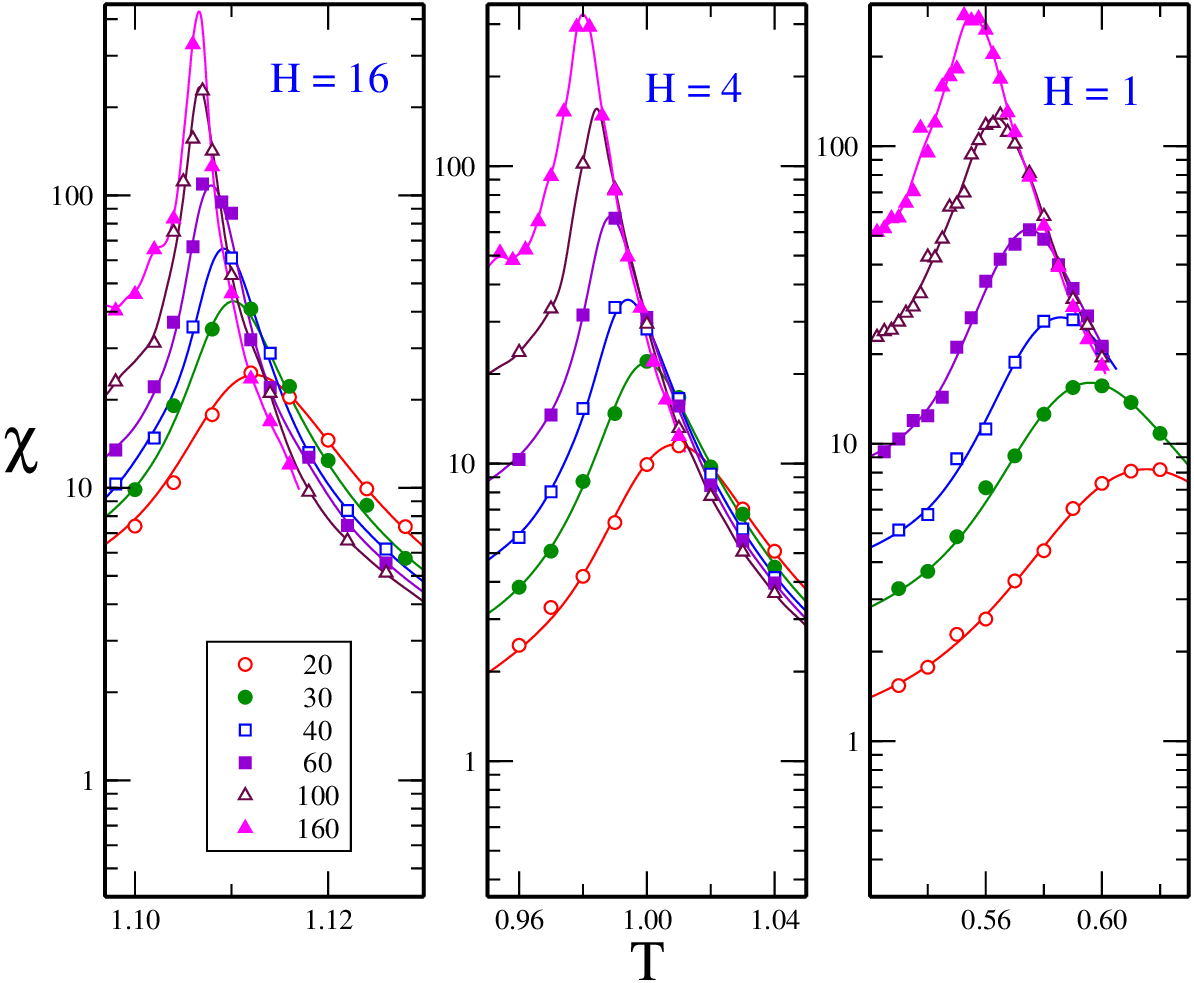}
\caption{(Color on line) Order parameter susceptibility, $\chi$ as a function of the temperature for different system widths,
$H=1,4,16$ and different systems sizes, $L$. Symbols and lines as in Fig.~\ref{fig.lambda}} 
\label{fig.susc}
\end{figure}

\begin{figure}
\includegraphics[width=90mm,clip=]{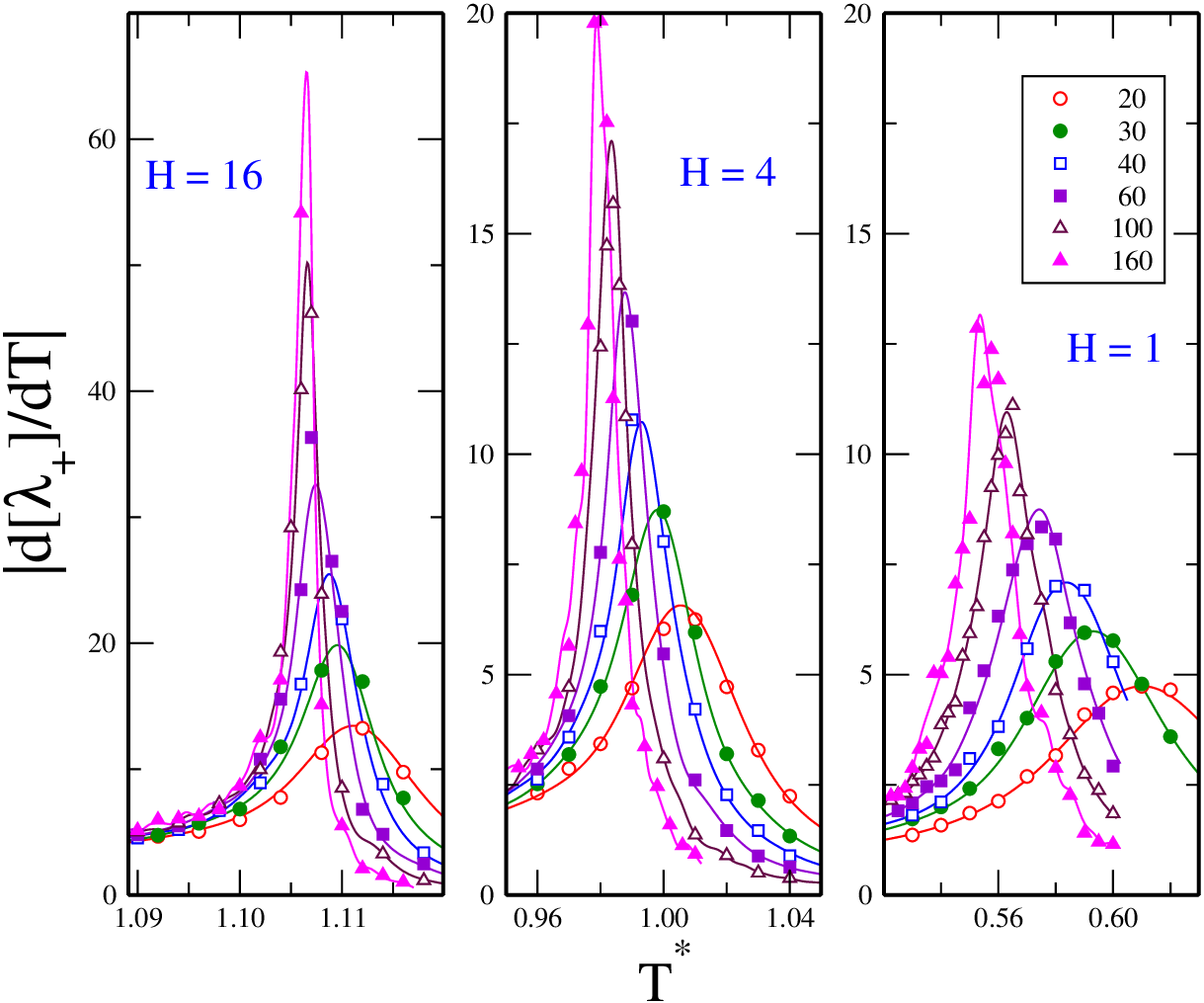}
\caption{(Color on line) Absolute value of the derivative of the nematic order parameter with respect to the temperature,
$|d\lambda_+/dT|$ as a function of the temperature for different system widths,
$H=1,4,16$ and different systems sizes, $L$. 
Symbols and lines as in Fig. \ref{fig.lambda}.}
\label{fig.dlambda}
\end{figure}

\begin{figure}
\includegraphics[width=90mm,clip=]{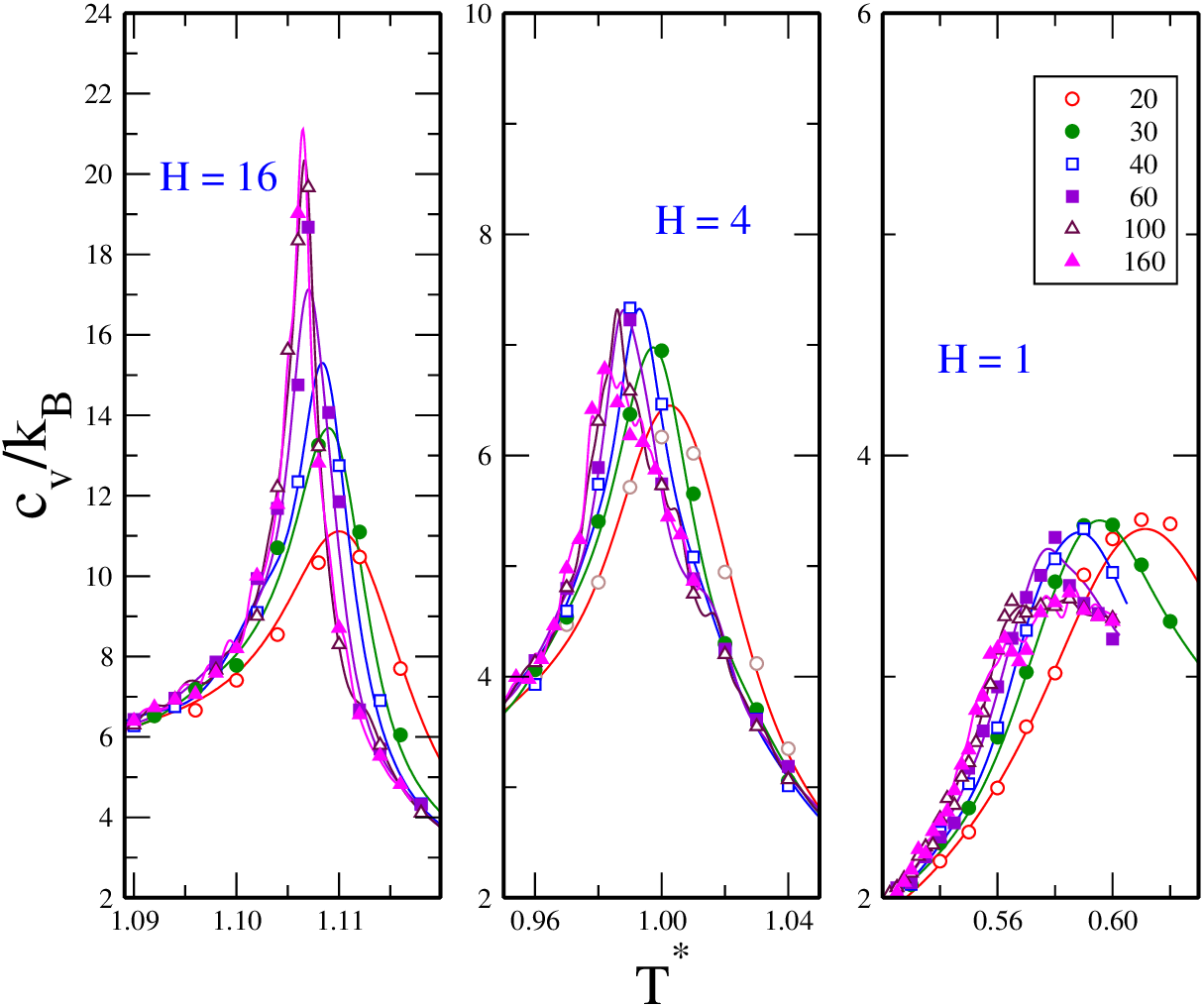}
\caption{(Color on line) Reduced excess heat capacity as a function of the temperature for different system widths,
$H=1,4,16$ and different systems sizes, $L$. 
Symbols and lines as in Fig. \ref{fig.lambda}.}
\label{fig.cv}
\end{figure}

In Figures \ref{fig.lambda}-\ref{fig.cv},
we depict the temperature dependence of $\lambda_+$, $\chi$,
$d\lambda_+/dT$ and $c_v$,  for different system sizes and pore widths. 

\begin{figure}
\includegraphics[width=90mm,clip=]{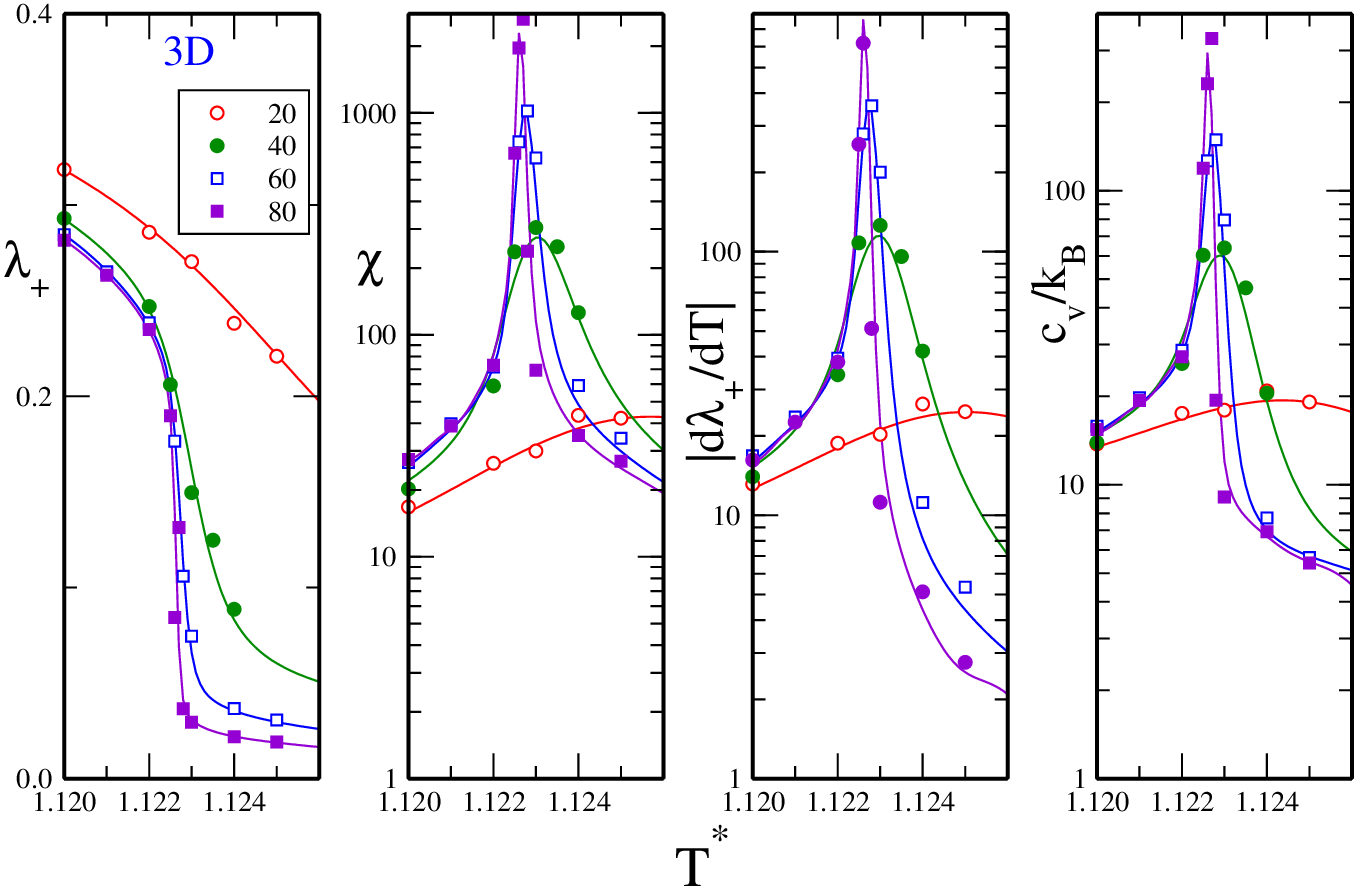}
\caption{(Color on line) Temperature and system size dependence of various properties
  of the 3D LL model (cubic cells without walls).
The length of the cell side, $L$, is shown  in the legends. Symbols
and lines as in previous figures.} 
\label{fig.3D}
\end{figure}

It can be seen that the dependence of these properties on $L$ is qualitatively
similar for the three slit widths considered in the figures. For a
given slit width, the susceptibility $\chi$ diverges 
with $L$ both at the temperature corresponding to the maximum and
below. The curves of $<\lambda_+>$ as a function of $T$ 
exhibit an inflection point, and the derivative of  $\lambda_+$ with respect to $T$ seems to 
diverge at a given critical temperature. 
It is to be stressed that the values of $\lambda_+$ below the apparent transition
temperature decrease with increasing sample sizes, in contrast with the expected
behavior from first and second order transitions.
On the other hand the heat capacity exhibits a maximum which does
not diverge with $L$. The behavior of all these properties around the
{\em transition} temperature resembles that of a topological BKT transition, and
it is clearly different from what one should expect in the presence
of a weak first order transition.
A second order transition might exhibit non-divergent heat capacity
curves (with negative
$\alpha$ exponent), but the decrease of $\lambda_+$ and divergence of
$\chi$ below the pseudo-critical $T_c(L,H)$, 
fit better into the picture of a continuous phase change which shares
a number of features with the continuous  BKT transition.
For the sake of comparison,  
in Figure \ref{fig.3D} we summarize the results of simulations for
unconfined systems using cubic boxes
of different sizes with full PBC. It seems
evident that the qualitative behavior 
of the confined system is quite different from that of the bulk, which
is known to present a weak first order isotropic-nematic transition 
\cite{Priezjev}. 

\begin{figure}
\includegraphics[width=90mm,clip=]{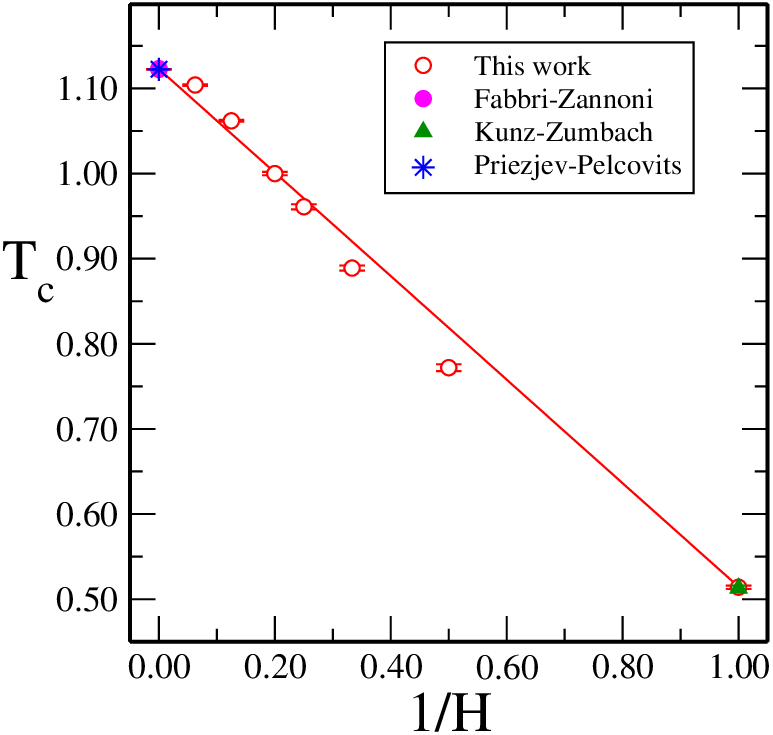}
\caption{(Color on line) Pore width dependence of the transition temperatures estimated
  using the scaling law (\ref{fit_bkt}). The result of a linear fit to
  $1/H$ is represented by a solid line. Values taken from
  Refs.\cite{Fabbri}, \cite{Priezjev}, and \cite{Kunz} are also
  included for comparison. Note that the bibliographic values and those
  of this work fall on  top of each other and can hardly be distinguished.}
\label{tbkt}
\end{figure}

Returning to the confined system, in table \ref{table}  we gather the results for the estimates of
its transition temperatures for different slit widths 
calculated using the two aforementioned definitions of the
pseudo-critical temperatures, and the two scaling laws. 
The results for $H=1$ agree with those reported in
Ref. \cite{Kunz}, but differ slightly from those reported in  
Ref.\cite{Mondal} using the scaling laws of {\it second order} transitions.
The results of the table show that the estimates of the transition
temperature are conditioned by the scaling law used in the  
extrapolation to the thermodynamic limit. However, the results hardly depend, within error bars,  on the particular definition
of the pseudo-critical temperature. The variation of the transition
temperature with $H$ is monotonic, and the transition temperatures
approach smoothly the bulk value as $H$ increases.  This is more
clearly seen in Figure \ref{tbkt}, where one can appreciate the
quasi-linear dependence of $T_c$ (as calculated from (\ref{fit_bkt}))
on $1/H$. This Kelvin-like scaling of the transition temperature leads
to an extrapolated value $\lim_{H\rightarrow\infty}T_c(H)= 1.123\pm
0.005$ that agrees rather well with the bulk value $T_c=1.1225(1)$,
which we have obtained using cubic systems with full PBC, and in
accordance with the results of Priezjev and
Pelcovits\cite{Priezjev}. 
Note, however that the Kelvin scaling only applies strictly to first
order phase transitions. 
In the case of second order transitions, correction terms must be
incorporated \cite{fisher1,fisher2,vink}. 
From our discussion it is clear that in our case a first order phase
transition is ruled 
out, so deviations from linearity could in principle be ascribed to
the continuous character of the transition. 
It is worth stressing that
$T_c$ estimates become
independent of 
the scaling relation used as $H$ increases. This is an indication,
that even if the transition can still be cast into the BKT-like type for
growing $H$, its scaling behavior is gradually switching to that of a
regular order-disorder transition. 

We also include in Table \ref{table} the estimates of the scaling exponent
for the maximum of the susceptibility, $\gamma/\nu$, which can be
drawn from the scaling relation:
\begin{equation}
\chi^{max}(L) \sim L^{\gamma/\nu}.
\end{equation}
The results of this exponent depend on the pore width, and 
constitute further evidence that no first-order
transition appears for the system sizes considered. One should expect
in this latter instance a scaling of the type $\chi^{max}(L) \sim H
L^2$, well away from the values obtained here for any width.
Incidentally, in the $H=1$ case the value is relatively close to the two
dimensional Ising critical exponent\cite{TCritPhen93}, $\gamma/\nu =
7/4$, and in agreement with the value reported by Kunz and
Zumbach\cite{Kunz} $\gamma/\nu=1.72\pm 0.05$. For larger slit widths,
the value of $\gamma/\nu$ decreases, what further deviates from the
limiting behavior of a first order transition
when $H\rightarrow\infty$ ($\gamma/\nu =2$). This implies that in the range
$0<H<\infty$ one should expect a non-monotonic behavior of
$\gamma/\nu$, as was already found in the confined HSMS
fluid\cite{Almarza1} and it is a clear indication that $H=16$ is still
far away from the first order transition limit. This situation is in
contrast with the results recently reported by Fish and
Vink\cite{Fish} for the generalized LL model with $p=8$. In this case
the angular interaction is much narrower than that of the simple LL
model and bears some resemblance with the $q$-state Potts
model\cite{Domany,Vink07} (with $q \propto \sqrt{p}$). Fish and Vink found that
$\gamma/\nu$ grows from 1.63 in the two dimensional limit approaching
$\gamma/\nu\rightarrow 2$ as the critical $H_c$ is reached and the
continuous transition develops into a first order transition. We
assume that as $p$ decreases $H_c$ increases (as observed in
Ref.~\cite{Fish} for $p\ge 8$), to the point that for $p=2$ the
determination of the critical thickness is
well beyond our present computational capabilities. On the other hand,
it is to be noticed the  fairly regular dependence of $T_c(H)$ on the
pore with.

\begin{table}
\begin{center}
\caption{Estimates of the isotropic-quasi-nematic transition temperature in the thermodynamic limit, using different
prescription of the pseudo-critical temperatures and scaling laws (See the text for details); and scaling exponents
$\gamma/\nu$ for the maxima of the susceptibility $\chi$. The transition temperature
in the
bulk system is $T_{N-I} =1.1225(1).$ Error bars across the table are shown between parentheses in units of the last figure and
correspond to a confidence level of 95 \%}
\label{table}
\begin{tabular}{l|rrrrrrr}
\hline
H                    & 1           &  2 &    3  &  4  & 5 & 8  & 16 \\
\hline 
$T_c$ [$\chi$, Eq.(\ref{fit_bkt})]           &   0.514(2) &  0.772(4) & 0.889(3) & 0.963(3) & 1.000(2) & 1.062(1) & 1.104(1)  \\
$T_c$ [$|d\lambda_+/dT|$, Eq.(\ref{fit_bkt})] &   0.508(4) &  0.764(7) & 0.889(5) & 0.956(4) & 0.999(3) & 1.062(2) & 1.104(1)  \\
$T_c$ [$\chi$, Eq.(\ref{fit_ct})]         &   0.536(3) &  0.782(3) & 0.905(3) & 0.973(3) & 1.010(2) & 1.067(1) & 1.105(1)  \\
$T_c$ [$|d\lambda_+/dT|$,Eq.(\ref{fit_ct})] &   0.531(4) &  0.786(5) & 0.906(4) & 0.969(3) & 1.008(2) & 1.067(1) & 1.105(1) \\
$\gamma/\nu$                                & 1.69(2) & 1.65(3) & 1.63(3) & 1.59(3) & 1.60(4) & 1.49(3) & 1.40(7) \\
\hline
\end{tabular}
\end{center}
\end{table}

\begin{figure}
\includegraphics[width=90mm,clip=]{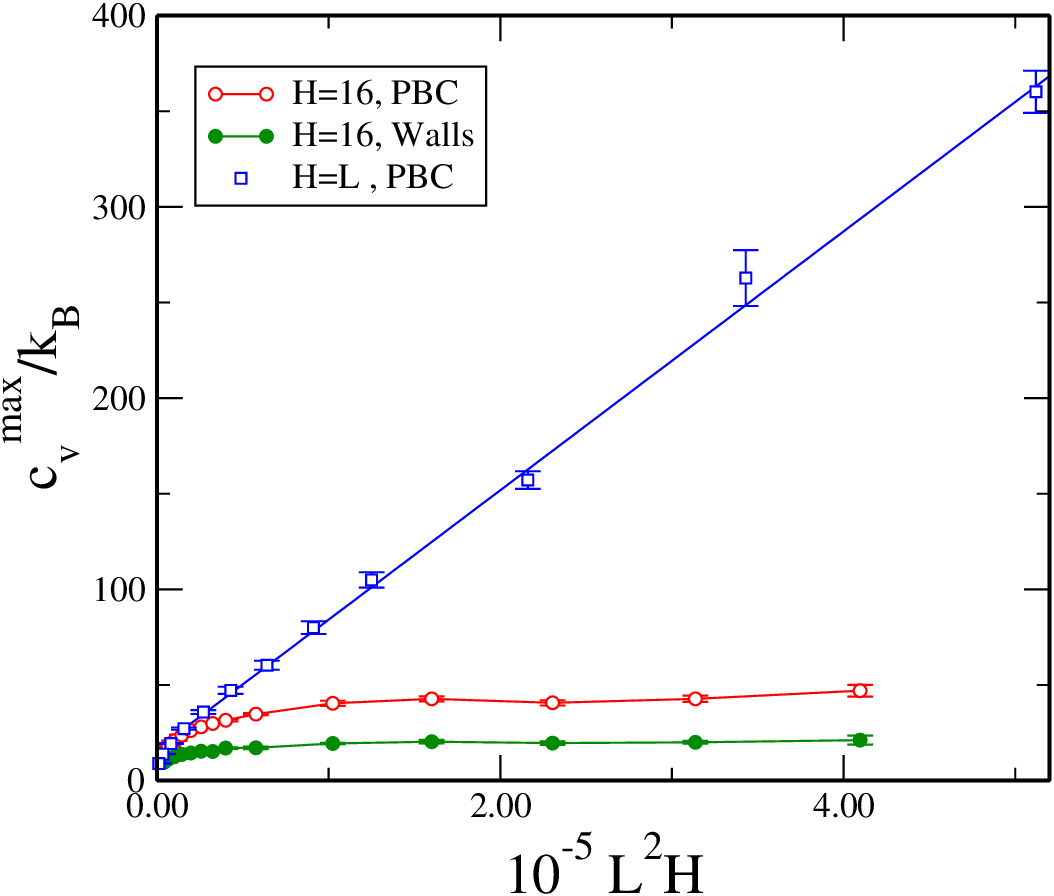}
\caption{(Color on line) Value of the maximum of  excess heat capacity per particle as a function of the system size for cubic systems (H=L)
with PBC (bulk LL), systems with H=16 and PBC (anisotropic LL), and
systems with H=16 confined between neutral walls. The line {\it connecting}
points for the bulk system corresponds to a fit of the results to the
equation $c_v(L)=a_0 + a_1 L^3 $} 
\label{fig.cvh16}
\end{figure}

From all this evidence, and in particular, from the size dependence of
the order parameter and the susceptibility, one can conclude
that the isotropic-quasi-nematic transitions found for all the confined systems
under scrutiny ($1 \le H \le 16$) do not fit in the picture of first or second order
phase changes, and share some resemblance with the continuous BKT transition.
The results also indicate that the possible critical thickness $H_c$ of the Lebwohl-Lasher model, if exists,
must appear for $H_c>16$. Therefore, the value of $H_c$ reported  in
Ref. [\onlinecite{cleaver}] is most likely underestimated.
It is possible to further analyze the effects of dimensionality on the
transition if the walls in the $z-$direction are replaced by PBC, but
still dealing with the $z$-direction on a different footing as compared
with the $x-$ and $y-$ directions. More precisely, we will consider a
series of systems of $L\times L\times H$ sites with PBC on all three
directions and for $H=2, 4,$ and 16.  This anisotropic LL model is
expected to enhance 
the correlations of the nematogen orientations  in the $z-$direction,
and eventually lead to the phase behavior of the bulk system as predicted by
theoretical arguments\cite{Shnidman}.  Interestingly,  we have found
that the anisotropic LL model with $H=2$, and $H=4$  also
exhibits BKT-like transitions similar to those of the corresponding
confined LL system occurring at slightly higher temperatures. Moreover,
the same behavior is found for $H=16$;  the system with PBC 
clearly shows a dependence of $c_v^{max}$ with $L$ inconsistent
with a first order transition. This can be appreciated in  Figure
\ref{fig.cvh16}, where we present the results for the value of the maximum of the excess heat capacity per molecule, $c_v^{max}$ for both systems.
It can be seen that for both confined, and PBC systems $c_v^{max}$ does not diverge. In the same Figure
we include the result for cubic systems $H=L$ with PBC (bulk LL
model); in this case the expected scaling behavior,  
$c_v^{max} \sim N $, of a first-order transition is observed.
From the results of $c_v^{max}$ it is possible to compute the latent heat, $\Delta E$ of
the transition\cite{landau_binder}:

\begin{equation}
c_v^{max}(N) = c_{\pm} + \frac{\left(\Delta E\right)^2 }{4 k_B T_c^2 } N;
\end{equation}
where $c_{\pm}$ is related with the specific heats of the two phases\cite{landau_binder}.
By fitting the results for $L \ge 25 $ we get $\Delta E/\epsilon = 0.0584 \pm 0.0013$.

\begin{figure}
\includegraphics[width=90mm,clip=]{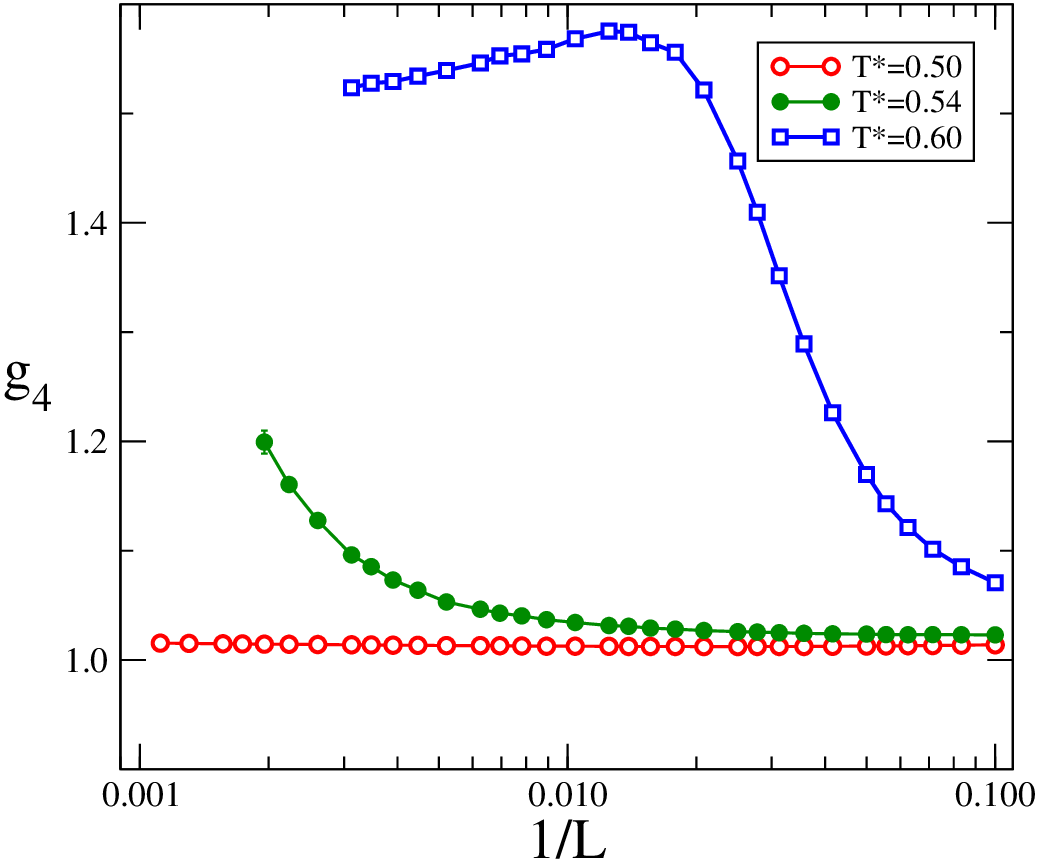}
\caption{(Color on line) Cumulant ratio $g_4=<\lambda_+^4>/<\lambda_+^2>^2$ as a function of the inverse
of the system size for $T^*=0.50$, $0.54$, and $0.50$}
\label{fig.g4a}
\end{figure}

\begin{figure}
\includegraphics[width=90mm,clip=]{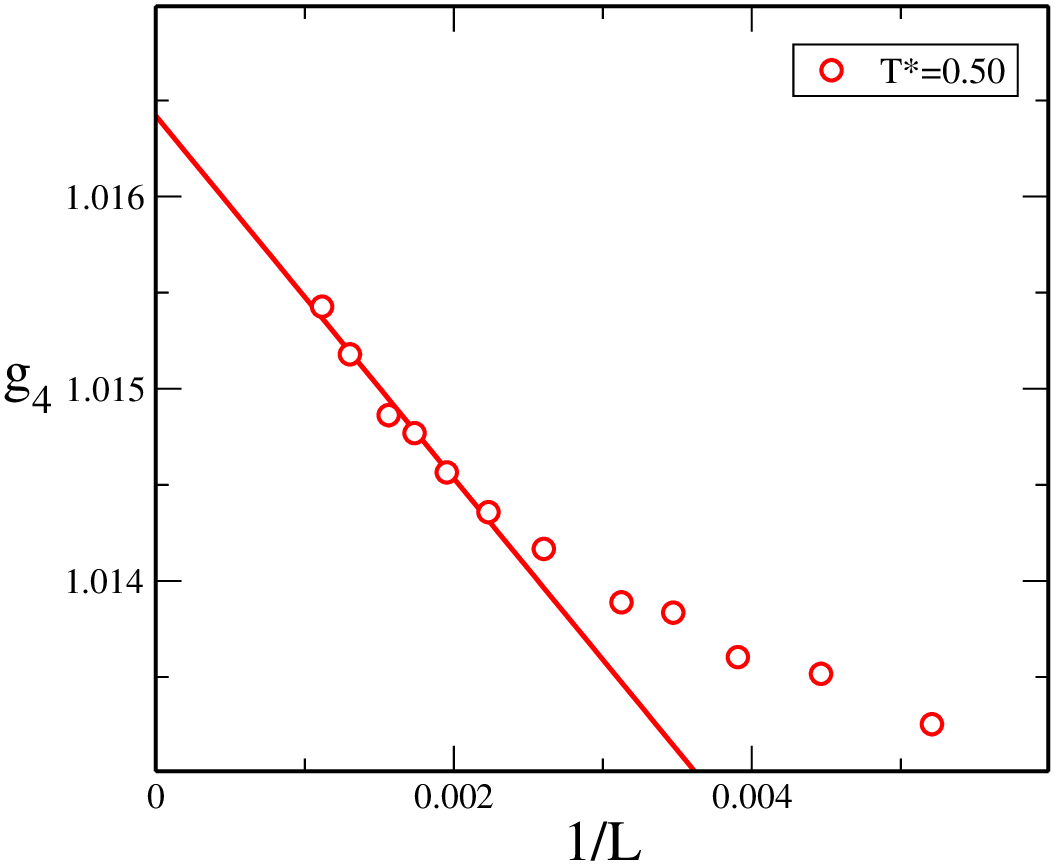}
\caption{(Color on line) Detail of the Cumulant ratio $g_4=<\lambda_+^4>/<\lambda_+^2>^2$ as a function of the inverse
of the system size for $T=0.50$}
\label{fig.g4b}
\end{figure}

In the discussion above, we have purposefully avoided an explicit
reference to the questions
recently raised by Paredes et al
\cite{Paredes,Farinas} as to the existence of a true BKT transition in
the PLL model.  After performing a finite-size scaling analysis
of the simulation results at temperatures around and below the estimates of the
transition temperature found in the literature, Paredes et al
\cite{Paredes,Farinas} conclude that the PLL lacks a true topological
transition. Using results for several of system sizes,  they infer
that the  $L$-dependence of the 
order parameter distribution for $T<T_{BKT}$ does not follow the expected scaling
 for a line of critical points. In particular,  they argue that the lack
 of crossing  
of the Binder cumulant \cite{binder,landau_binder,weber} curves for
different system sizes at  
a fixed temperature  is a strong evidence of the absence
of quasi long range order in the PLL model. 
In order to gain some extra insight into this problem,  we have
performed series of simulations 
with a broad range of system sizes at three temperatures:
$T^*=0.50$ (slightly below the range of our
$T_c$ estimates), $T^*=0.54$ (slightly above), and $T^*=0.60$.  
The so-called Binder cumulant, $g_4$ can be defined as\cite{weber}:
\begin{equation}
g_4  = \frac{< \lambda_+^4> }{<\lambda_+^2>^2 }.
\end{equation}
It is well established\cite{landau_binder} that for
second order transitions at the critical temperature $g_4$ reaches (for large system sizes)
a critical value (different from those corresponding to ordered and disordered phases) which becomes
independent on the system size, $g_4(L,T_c)=g_4^{(c)}$. According to the usual description 
of the topological transitions, 
below $T_{BKT}$ there should be a line of critical points, and therefore at a fixed temperature $g_4(L;T)$ should
approach a critical value $g_4^{c}(T)$ as $L \rightarrow \infty$. This
value must be different from those of the
isotropic ($g_4\approx 3/2$)  and nematic ($g_4\approx 1$)  phases.
From this point of view one can expect that plotting $g_4$ as a function
of $T$, the curves with different values of $L$ should merge for $T \le T_{BKT}$. Of course, finite
size effects could eventually lead to a small degree of {\it crossing} (See Ref. [\onlinecite{Paredes}]).
Therefore, from our point of view the absence  of crossing between the $g_4(T)$ curves with
different values of $L$ must not be regarded as a signature of lack of criticality. In figures
\ref{fig.g4a} and \ref{fig.g4b} we show the results of $g_4$ as a function of the system size 
for $T^*=0.50$ and $T^*=0.54$, and $T=0.60$.
The results seem to be compatible with the presence of a BKT-like transition in the PLL model.
 For large system sizes, at $T=0.54$, and $T=0.60$ $g_4(L)$ seems to
 approach  to the expected value 
for the isotropic phase ($g_4 \approx 3/2$),  whereas for $T=0.50$ (with system sizes up to $L=896$)
the values of $g_4(L)$ apparently converge towards a critical value as $L \rightarrow \infty$.

Another point raised by Paredes et al.\cite{Paredes} concerns the
apparent violation of the hyperscaling relation of the critical
exponents, $2\beta/\nu + \gamma/\nu = d$ (where $d$ is the
space dimensionality). In the case of a BKT transition, the exponent
$\nu$ is not defined, but the exponent ratios can still be
calculated\cite{Kost}. According to Ref.~\cite{Paredes} in the case of the
PLL this relation is only fulfilled within a 3\% accuracy, one order
of magnitude less than in the case of the XY-model. In our case,
calculations carried out at $T^* = 0.50$ (below the transition
temperature) also indicate deviations
around 5\%, somewhat larger than the statistical
uncertainties. Interestingly, previous calculations performed at the
transition temperature for the 
continuum HSMS model \cite{Lomba3} agree  with the hyperscaling
behavior  within a 0.7\% error. Moreover, using an alternative
definition of the susceptibility for temperatures above
$T_c(H,L)$\cite{landau_binder,Peczak,Chamati}
\[ \chi = N <\lambda_+^2>/k_B T,\]
 we found that the hyperscaling relation is appropriately fulfilled.

\begin{figure}
\includegraphics[width=90mm,clip=]{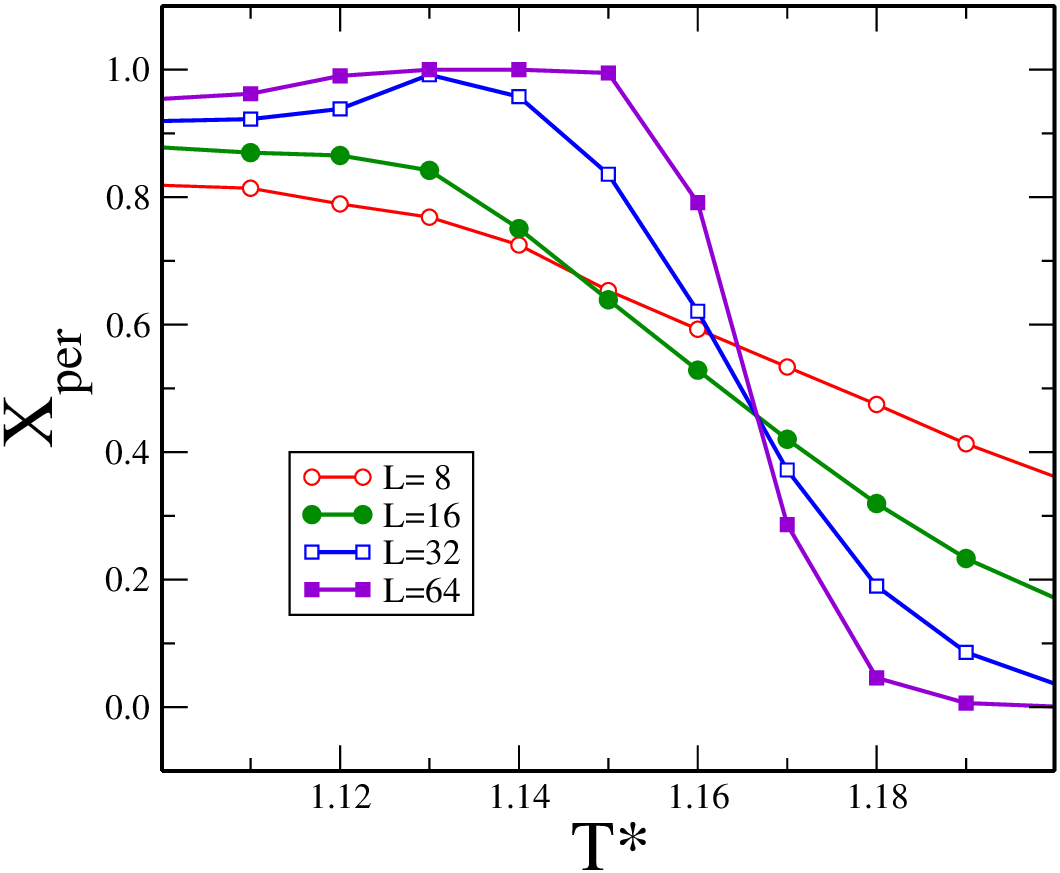}
\caption{(Color on line) Probability of finding  cluster percolation in, at least, one direction
in the simulation of the three-dimensional Lebwohl-Lasher model as a function of the
temperature. Different curves represent results for different system sizes.}
\label{fig.per3d}
\end{figure}

\begin{figure}
\includegraphics[width=90mm,clip=]{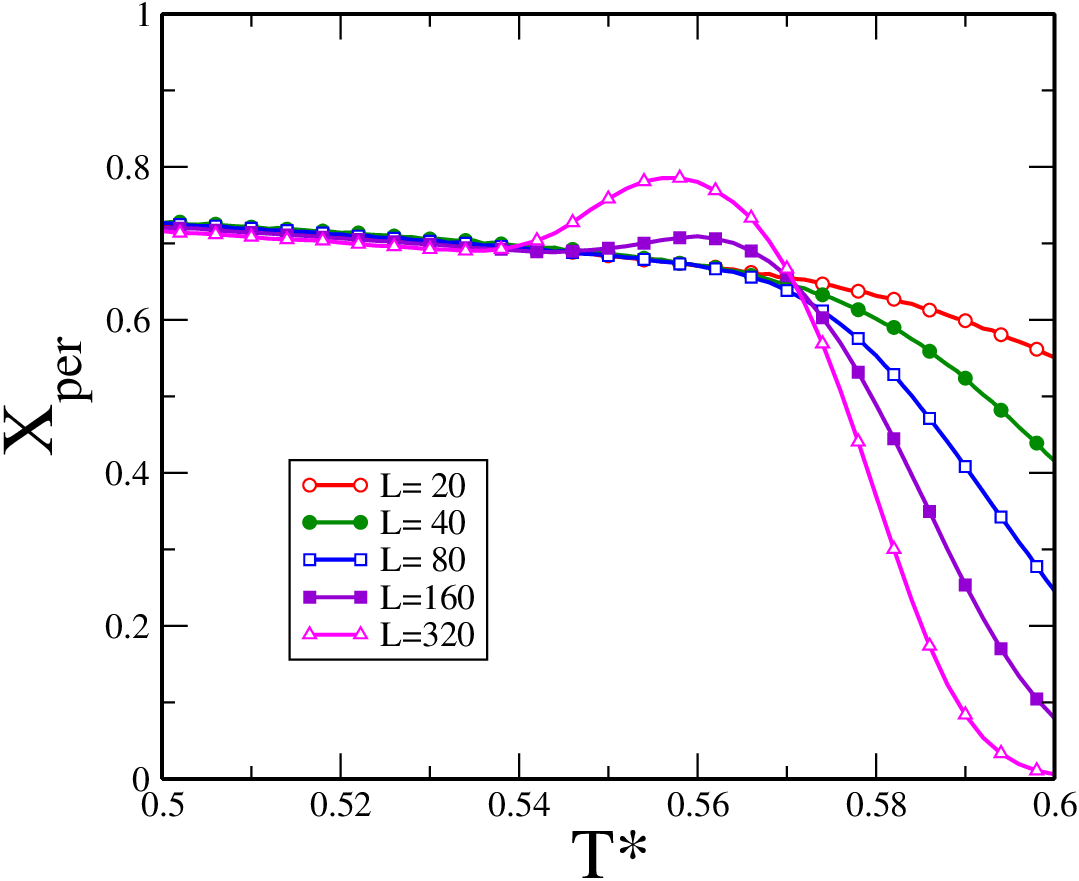}
\caption{(Color on line) Probability of finding cluster percolation in, at least, one direction,
in the simulation of the planar Lebwohl-Lasher model as a function of the
temperature. Different curves represent results for different system sizes. }
\label{fig.per2d}
\end{figure}

Some additional information can be obtained from an analysis of the
percolation of the clusters constructed by the simulation 
algorithm as a function of the temperature, so as to evaluate the
degree of correlation between the particle orientations within
the simulated samples.  
Let us recall that the Swendsen-Wang-like (SW) algorithm applied in this work belong to the class
of rejection-free cluster methods, and for some simple systems 
the temperature at which the cluster percolation occurs corresponds to that of the phase transition.
This property was used by Tomita and Okabe\cite{Tomita} to locate the BKT transitions of
two-dimensional $XY$ and Potts-Clock models in two dimensions. 
For the PLL model we  have carried out 
multi-temperature simulations using the {\em single tempering} algorithm of Zhang and Ma\cite{zhang}.
In Figure \ref{fig.per3d} we present the results of the percolation probability,
$X_{per}$, defined as the fraction of configurations containing at least one percolating cluster, for
the 3D LL model with PBC.
It can be seen that  the percolation threshold appears at a
temperature slightly above the nematic-isotropic 
transition temperature. In addition. $X_{per}(L,T)$ shows a non-monotonic behavior with $T$ for large system sizes
at temperatures close to the thermodynamic transition.
The behavior of $X_{per}(L,T)$ is qualitatively similar 
for the PLL model (See Fig. \ref{fig.per2d}); and at temperatures close to
the $T_{BKT}$ estimates the curves for different system sizes show a clear tendency
to merge. The crossing of the curves for different system
sizes observed for large system size
seem to indicate that the aforementioned merging
 is not just a consequence of correlations induced by the periodic
 boundary conditions. Moreover, in a similar percolation analysis
 carried out by us for the planar Heisenberg model (3D spins on a plane),
 the  $X_{per}(L,T)$ curves for different system sizes do not exhibit
 any crossing at finite temperatures, but seemingly merge as
 $T\rightarrow 0$. This supports the general view\cite{blotte} that
 the 2D Heisenberg model does not have a phase transition at $T >0$,
 and underlines the essentially different phase behavior of the PLL
 model. In the thermodynamic limit, Figure \ref{fig.per3d} seems to
 indicate that in the case of the 3D LL model, there should be an
 abrupt change from the non-percolating state ($X_{per}=0$) to a fully
 percolating state ($X_{per}=1$) at a finite temperature, which fits
 into the picture of a first order transition between the isotropic
 and a truly nematic phase. In contrast, in Figure  \ref{fig.per2d},
 one finds that in the PLL model, at least for the system sizes here
 considered, the fully percolating state is only reached at
 $T=0$. Note, that by construction, the SW cluster algorithm may yield
 $X_{per} <1$ for finite temperatures even in the case of truly
orientationaly ordered states. The presence of the maximum after the
first crossing (occurring both in the 3D LL and PLL models for finite
sizes, but seemingly not present in the 3D Heisenberg system) might
then well be an effect of the cluster algorithm. On the other hand,
analysing the size dependence of the curves plotted in Figure
\ref{fig.per2d}, one is tempted to assume that the maxima will continue
to grow and shift to lower T as L increases, until finally $X_{per}=1$
is reached for a given sample size. Whether this is really the case,
and if so, $X_{per}=1$ is reached at $T>0$ or not, cannot be assessed
at present using reasonable computer resources.

In any case, we believe that the percolation analysis sketched above confirms that the PLL model indeed
presents a phase transition. It might be the case, that we are not dealing 
here with a strict BKT transition, if one takes into account the
previous discussion on the hyperscaling relation, but its phenomenology is
closely related to that of the BKT transition.  On the other hand, the other
anomalies in the model's scaling behavior  found by Paredes et al.
\cite{Paredes} could be ascribed to finite size effects. 
To conclude this discussion about the likelihood of a topological
transition for the PLL model, 
it is worth to compare the phase behavior of the three-dimensional XY and LL models. The three-dimensional
XY model presents a continuous transition \cite{campostrini} without a divergence in the specific heat,
whereas the three-dimensional LL model exhibits a first order transition. Taking as a reference the critical
behavior of Potts models \cite{wu}, we would not expect in principle that the PLL model had a {\em weaker}
transition than that of the two-dimensional XY model.

In summary, we have studied 
the order-disorder transition for the confined LL model by means of
Monte Carlo simulation and 
finite-size scaling analysis. Our results indicate that the critical
pore width signaling 
the crossover between bulk and 2D behavior must be larger than the
values indicated by previous simulations. The need for a reliable
finite-size scaling 
analysis on systems with larger widths, which would imply simulations
for much larger 
systems hampers the estimation of $H_c$. In addition, our results for slit-like
systems with full PBC (anisotropic LL model) suggest that if $H_c$ has
a finite value, it will likely be much larger 
than $H=16$. Moreover, the fact that the critical exponent relation
$\gamma/\nu$ for the pore widths considered does  not yet show any
trend to converge towards the expected behavior in a first order
transition, is a further indication that we are very likely away from
the critical thickness. This might fit into the picture drawn by Telo
da Gama and Tarazona\cite{telodagama}, who suggest that one should expect
$H_c\rightarrow\infty$.  However, in Ref. \onlinecite{telodagama} it is argued
that spin-waves would destroy the ordered phase for any finite $H$, but no BKT transition would occur.
Our findings suggest that an order-disorder phase transition with some BKT-like features does
indeed take place, as the divergence of the susceptibility, its size dependence, the crossover
of percolation curves and the size dependence of the order parameter seem to evidence.
 It is worth pointing out that the situation
depicted here is in marked contrast with the abrupt switch from
continuous 2D melting behavior to discontinuous first order melting
which has been argued to occur in hard sphere colloidal models when
going from monolayer to bilayer systems\cite{Xu}.

\acknowledgments
The authors gratefully acknowledge the support from the Direcci\'on
General de Investigaci\'on Cient\'{\i}fica  y T\'ecnica under Grant
No. MAT2007-65711-C04-04 and from the Direcci\'on General de
Universidades e Investigaci\'on de la Comunidad de Madrid under
Grant No. S2009/ESP-1691 and Program MODELICO-CM.


\end{document}